%% file: TSP2010.tex
\documentclass[journal]{IEEEtran}
\usepackage{amsmath}
\usepackage{amsfonts}
\usepackage{amsthm}
\usepackage{amssymb}
\usepackage{cite}
\usepackage{url}
\usepackage{subfigure}
\usepackage{graphicx}
\usepackage{color}

\include{Defn}

\begin{document}

\title{Robust Shrinkage Estimation of High-dimensional Covariance Matrices}
\author{Yilun~Chen,~\IEEEmembership{Student~Member,~IEEE,} Ami~Wiesel,~\IEEEmembership{Member,~IEEE,} and Alfred~O.~Hero~III,~\IEEEmembership{Fellow,~IEEE}
\thanks{Y. Chen and A. O. Hero are with the Department
of Electrical Engineering and Computer Science, University of
Michigan, Ann Arbor, MI 48109, USA. Tel: 1-734-763-0564. Fax:
1-734-763-8041.
Emails: \{yilun, hero\}@umich.edu.}
\thanks{A. Wiesel is with the The Rachel and Selim Benin School of 
Computer Science and Engineering at the Hebrew University of Jerusalem, 
91904 Jerusalem, Israel. Tel: 972-2-6584933. Email: ami.wiesel@huji.ac.il.}
\thanks{This work was partially supported by AFOSR, grant number FA9550-06-1-0324. The work of A. Wiesel was supported by
a Marie Curie Outgoing International Fellowship within the 7th
European Community Framework Programme. Parts of this work were
presented at the 2010 IEEE Workshop on Sensor Array and
Multichannel Signal Processing (SAM).}
}
\maketitle

\begin{abstract}
We address high dimensional covariance estimation for elliptical
distributed samples, which are also known as spherically invariant
random vectors (SIRV) or compound-Gaussian processes. Specifically
we consider shrinkage methods that are suitable for high
dimensional problems with a small number of samples (large $p$
small $n$). We start from a classical robust covariance estimator
[Tyler(1987)], which is distribution-free within the family of
elliptical distribution but inapplicable when $n<p$. Using a
shrinkage coefficient, we regularize Tyler's fixed point
iterations. We prove that, {for all $n$ and $p$,} the proposed
fixed point iterations converge to a unique limit regardless of
the initial condition. Next, we propose a simple, closed-form and
data dependent choice for the shrinkage coefficient, which is
based on a minimum mean squared error framework. Simulations
demonstrate that the proposed method achieves low estimation error
and is robust to heavy-tailed samples. Finally, as a real world
application we demonstrate the performance of the proposed
technique in the context of activity/intrusion detection using a
wireless sensor network.
\end{abstract}




\section{Introduction}
\label{sec:intro} Estimating a covariance matrix (or a dispersion
matrix) is a fundamental problem in statistical signal processing.
Many techniques for detection and estimation rely on accurate
estimation of the true covariance. In recent years, estimating a
high dimensional $p \times p$ covariance matrix under small sample
size $n$ has attracted considerable attention. In these ``large
$p$ small $n$'' problems, the classical sample covariance suffers
from a systematically distorted eigen-structure \cite{Jonestone},
and improved estimators are required.

Much effort has been devoted to high-dimensional covariance
estimation, which use Steinian shrinkage
\cite{Stein,Ledoit2004,Chen} or other types of regularized methods
such as \cite{Bickel-Levina06, graph-lasso}. However, most of the
high-dimensional estimators assume Gaussian distributed samples.
This limits their usage in many important applications involving
non-Gaussian and heavy-tailed samples. One exception is the
Ledoit-Wolf estimator \cite{Ledoit2004}, where the authors shrink
the sample covariance towards a scaled identity matrix and
proposed a shrinkage coefficient which is asymptotically optimal
for any distribution. However, as the Ledoit-Wolf estimator
operates on the sample covariance, it is inappropriate for heavy
tailed non-Gaussian distributions. On the other hand, traditional
robust covariance estimators \cite{Huber81, Tyler87, Rousseeuw85}
designed for non-Gaussian samples generally require $n \gg p$ and
are not suitable for ``large $p$ small $n$'' problems. Therefore,
the goal of our work is to develop a covariance estimator for
problems that are both high dimensional and non-Gaussian. In this
paper, we model the samples using the elliptical distribution
\cite{Kelker}, which is also referred to as the spherically
invariant random vector model (SIRV) \cite{Yao,Yao04} or the
compound-Gaussian process model \cite{Conte1995}. As a flexible
and popular alternative, the elliptical family encompasses a large
number of important distributions such as Gaussian distribution,
the multivariate Cauchy distribution, the multivariate exponential
distribution, the multivariate Student-T distribution, the
K-distribution and the Weibull distribution. The capability of
modelling heavy-tails makes the elliptical distribution appealing
in signal processing and related fields. Typical applications
include radar detection \cite{MIT,Conte1995, Rangaswamy05,Wang},
speech signal processing \cite{Brehm87}, remote sensing
\cite{Vasile}, wireless fading channels modelling \cite{Yao04},
financial engineering \cite{Frahm2004} and so forth.

A well-studied covariance estimator for elliptical distributions
is the ML estimator based on normalized samples
\cite{Tyler87,Gini02, Conte02}. The estimator is derived as the
solution to a fixed point equation by using fixed point
iterations. It is distribution-free within the class of elliptical
distributions and its performance advantages are well known in the
$n\gg p$ regime. However, it is not suitable for the ``large $p$
small $n$" setting. Indeed, when $n < p$, the ML estimator as
defined does not even exist. To avoid this problem the authors of
\cite{Abrmovich} propose an iterative regularized ML estimator
that employs diagonal loading and uses a heuristic procedure for
selecting the regularization parameter. While they did not
establish convergence and uniqueness \cite{Abrmovich} they
empirically demonstrated that their algorithm has superior
performance in the context of a radar application. Our approach is
similar to \cite{Abrmovich} but we propose a systematic procedure
of selecting the regularization parameter and establish
convergence and uniqueness of the resultant iterative estimator.
Specifically, we consider a shrinkage estimator that regularizes
the fixed point iterations. For a fixed shrinkage coefficient, we
prove that the regularized fixed iterations converge to a unique
solution {for all $n$ and $p$}, regardless of the initial
condition. Next, following Ledoit-Wolf \cite{Ledoit2004}, we
provide a simple closed-form expression for the shrinkage
coefficient, based on minimizing mean-squared-error. The resultant
coefficient is a function of the unknown true covariance and
cannot be implemented in practice. Instead, we develop a
data-dependent ``plug-in" estimator approximation. Simulation
results demonstrate that our estimator achieves superior
performance for samples distributed within the elliptical family.
Furthermore, for the case that the samples are truly Gaussian, we
report very little performance degradation with respect to the
shrinkage estimators designed specifically for Gaussian samples
\cite{Chen}.

As a real world application we demonstrate the proposed estimator
for activity/intrusion detection using an active wireless sensor
network. We show that the measured data exhibit strong
non-Gaussian behavior and demonstrate significant performance
advantages of the proposed robust covariance estimator when used
in a covariance-based anomaly detection algorithm.

The paper is organized as follows. Section II provides a brief
review of elliptical distributions and of Tyler's covariance
estimation method. The regularized covariance estimator is
introduced and derived in Section III. We provide simulations and
experimental results in Section IV and Section V, respectively.
Section VI summarizes our principal conclusions. The proof of
theorems and lemmas are provided in the Appendix.

\emph{Notations}: In the following, we depict vectors in lowercase
boldface letters and matrices in uppercase boldface letters.
$(\cdot)^T$ denotes the transpose operator. $\tr(\cdot)$ and
$\det(\cdot)$ are the trace and the determinant of a matrix,
respectively.



\section{ML covariance estimation for elliptical distributions}
\subsection{Elliptical distribution}
Let $\bx$ be a $p \times 1$ zero-mean random vector generated by
the following model
\begin{equation}
\bx = \nu \bu, \label{eq:1}
\end{equation}
where $\nu$ is a positive random variable and $\bu$ is a $p \times
1$ zero-mean, jointly Gaussian random vector with positive
definite covariance $\bbsg$. We assume that $\nu$ and $\bu$ are
statistically independent. The resulting random vector $\bx$ is
elliptically distributed and its probability density function
(pdf) can be expressed by
\begin{equation}
\label{eq:ellp_pdf} p(\bx) = \phi \bl \bx^T\bbsg^{-1}\bx\br,
\end{equation}
where $\phi(\cdot)$ is the characteristic function (Definition 2,
pp. 5, \cite{Frahm2004}) related to the pdf of $\nu$. The
elliptical family encompasses many useful distributions in signal
processing and related fields and includes: the Gaussian
distribution itself, the K distribution, the Weibull distribution
and many others. As stated above, elliptically distributed samples
are also referred to as Spherically Invariant Random Vectors
(SIRV) or compound Gaussian processes in signal processing.

\subsection{ML estimation}
Let $\blc \bx_i\brc_{i=1}^n$ be a set of $n$ independent and
identically distributed (i.i.d.) samples drawn according to
(\ref{eq:1}). The problem is to estimate the covariance
(dispersion) matrix $\bbsg$ from $\blc \bx_i\brc_{i=1}^n$. The
model (\ref{eq:1}) is invariant to scaling of the covariance
matrix $\bbsg$ of $\bu$. Therefore, without loss of generality, we
assume that the covariance matrix is trace-normalized in the sense
that $\tr(\bbsg)=p$.

The commonly used sample covariance, defined as
\begin{equation}
\hs = \frac{1}{n} \sum_{i=1}^n \bx_i \bx_i^T, \label{eq:3}
\end{equation}
is known to be a poor estimator of $\bbsg$, especially when the
samples are high-dimensional (large $p$) and/or heavy-tailed.
Tyler's method \cite{Tyler87} addresses this problem by working
with the normalized samples:
\begin{equation}
\bs_i = \frac{\bx_i}{\|\bx_i\|_2} = \frac{\bu_i}{\|\bu_i\|_2},
\label{eq:5}
\end{equation}
for which the term $\nu$ in (\ref{eq:1}) drops out. The pdf of
$\bs_i$ is given by \cite{Frahm2004}
\begin{equation}
p(\bs_i;\bbsg) = \frac{\Gamma(p/2)}{2\pi^{p/2}} \cdot
\sqrt{\det(\bbsg^{-1})} \cdot \bl {\bs_i^T\bbsg^{-1}\bs_i}
\br^{-p/2}. \label{eq:8}
\end{equation}
Taking the derivative and equating to zero, the maximum likelihood
estimator based on $\{\bs_i\}_{i=1}^n$ is the solution to
\begin{equation}
\bbsg = \frac{p}{n} \cdot \sum_{i=1}^n \frac{\bs_i
\bs_i^T}{\bs_i^T \bbsg^{-1} \bs_i}. \label{eq:4}
\end{equation}

When $n \ge p$, the ML estimator can be found using the following
fixed point iterations:
\begin{equation}
\hsg_{j+1} = \frac{p}{n} \cdot \sum_{i=1}^n \frac{\bs_i
\bs_i^T}{\bs_i^T \hsg_j^{-1} \bs_i}, \label{eq:7}
\end{equation}
where the initial $\hsg_{0}$ is usually set to the identity
matrix. {Assuming that $n \ge p$ and that any $p$ samples out of
$\{\bs_i\}_{i=1}^n$ are linearly independent with probability
one,} it can be shown that the iteration process in (\ref{eq:7})
converges and that the limiting value is unique up to constant
scale, which does not depend on the initial value of $\hsg_{0}$.
In practice, a final normalization step is needed, which ensures
that the iteration limit $\hsg_{\infty}$ satisfies
$\tr(\hsg_{\infty}) = p$.



The ML estimate corresponds to the Huber-type M-estimator and has
many good properties when $n \gg p$, such as asymptotic normality
and strong consistency. Furthermore, it has been pointed out
\cite{Tyler87} that the ML estimate (\ref{eq:7}) is the ``most
robust" covariance estimator in the class of elliptical
distributions in the sense of minimizing the maximum asymptotic
variance. We note that (\ref{eq:7}) can be also motivated from
other approaches as proposed in \cite{Gini02, Conte02}.

\section{Robust shrinkage covariance estimation}
Here we extend Tyler's method to the high dimensional setting
using shrinkage regularization. It is easy to see that there is no
solution to (\ref{eq:4}) when $n<p$ (its left-hand-side is full
rank whereas its right-hand-side of is rank deficient). This
motivates us to develop a regularized covariance estimator for
elliptical samples. Following \cite{Ledoit2004, Chen}, we propose
to regularize the fixed point iterations as
\begin{align}
\label{eq:iterations} \tsg_{j+1} & = (1-\rho) \frac{p}{n}
\sum_{i=1}^n \frac{\bs_i
\bs_i^T}{\bs_i^T \hsg_{j}^{-1} \bs_i} + \rho \bbi,\\
\label{eq:normalize} \hsg_{j+1} & =
\frac{\tsg_{j+1}}{\tr(\tsg_{j+1})/p},
\end{align}
where $\rho$ is the so-called shrinkage coefficient, which is a
constant between 0 and 1. When $\rho=0$ and $n \ge p$ the proposed
shrinkage estimator reduces to Tyler's unbiased method in
(\ref{eq:4}) and when $\rho = 1$ the estimator reduces to the
trivial uncorrelated case yielding a scaled identity matrix. The
term $\rho \bbi$ ensures that $\hsg_{j+1}$ is always
well-conditioned and thus allows continuation of the iterative
process without the need for restarts. Therefore, the proposed
iteration can be applied to high dimensional estimation problems.
We emphasize that the normalization (\ref{eq:normalize}) is
important and necessary for convergence. We establish provable
convergence and uniqueness of the limit in the following theorem.



\begin{thm}
\label{thm:2} Let $0<\rho<1$ be a shrinkage coefficient. Then, the
fixed point iterations in (\ref{eq:iterations}) and
(\ref{eq:normalize}) converge to a unique limit for any positive
definite initial matrix $\hsg_0$.
\end{thm}


The proof of Theorem \ref{thm:2} follows directly from the concave
Perron-Frobenius theory \cite{Krause94} and is provided in the
Appendix. We note that the regularization presented in
(\ref{eq:iterations}) {and (\ref{eq:normalize})} is similar to
diagonal loading \cite{Abrmovich}. However, unlike the diagonal
loading approach of \cite{Abrmovich}, the proposed shrinkage
approach provides a systematic way to choose the regularization
parameter $\rho$, discussed in the next section.

\subsection{Choosing the shrinkage coefficient}
We now turn to the problem of choosing a good, data-dependent,
shrinkage coefficient $\rho$, as as an alternative to
cross-validation schemes which incur intensive computational
costs. As in Ledoit-Wolf \cite{Ledoit2004}, we begin by assuming
we ``know" the true covariance $\bbsg$. Then we define the
following clairvoyant ``estimator":
\begin{equation}
\label{eq:tsg_def} \widetilde{\bbsg}(\rho) =
(1-\rho)\frac{p}{n}\sum_{i=1}^n \frac{\bs_i \bs_i^T}{\bs_i^T
\bbsg^{-1} \bs_i} + \rho \bbi,
\end{equation}
where the coefficient $\rho$ is chosen to minimize the minimum
mean-squared error:
\begin{equation}
\rho_O=\arg\min_\rho
E\blc\l\|\widetilde{\bbsg}(\rho)-\bbsg\r\|_F^2\brc.
\label{eq:mmse}
\end{equation}
The following theorem shows that there is a closed-form solution
to the problem (\ref{eq:mmse}), which we refer to as the ``oracle"
coefficient.
\begin{thm}
\label{thm:rho} For i.i.d. elliptical distributed samples the
solution to (\ref{eq:mmse}) is
\begin{equation}
\label{eq:cov_oracle} \rho_O =
\frac{p^2+(1-2/p)\tr(\bbsg^2)}{(p^2-np-2n) +
(n+1+2(n-1)/p)\tr(\bbsg^2)},
\end{equation}
under the condition $\tr(\bbsg) =p$.
\end{thm}
The proof of Theorem \ref{thm:rho} requires the calculation of the
fourth moments of an isotropically distributed random vector
\cite{Marzetta, Hassibi, Eldar_} and is provided in the Appendix.

The oracle coefficient cannot be implemented since $\rho_O$ is a
function of the unknown true covariance $\bbsg$. Therefore, we
propose a plug-in estimate for $\rho_O$:
\begin{equation}
\label{eq:plug_in} \hat \rho =
\frac{p^2+(1-2/p)\tr(\hm^2)}{(p^2-np-2n) +
(n+1+2(n-1)/p)\tr(\hm^2)},
\end{equation}
where $\hm$ can be any consistent estimator of $\bbsg$, \eg the
trace-normalized Ledoit-Wolf estimator. Another appealing
candidate for plug-in is the (trace-normalized) normalized sample
covariance $\hr$ \cite{Conte1994} defined by:
\begin{equation}
\hr = \frac{{p}}{n}\sum_{i=1}^n \bs_i \bs_i^T.
\end{equation}
We note that the only requirement on the covariance estimator
$\hm$ is that it provide a good approximation to {$\tr(\bbsg^2)$}.
It does not have to be well-conditioned nor does it have to be an
accurate estimator of the true covariance matrix $\bbsg$.

By using the plug-in estimate $\hat\rho$ in place of $\rho$, the
robust shrinkage estimator is computed via the fixed point
iteration in (\ref{eq:iterations}) {and (\ref{eq:normalize})}.




\section{Numerical simulation}
In this section we use simulations to demonstrate the superior
performance of the proposed shrinkage approach. First we show that
our estimator outperforms other estimators for the case of
heavy-tailed samples generated by a multivariate Student-T
distribution, where $\nu$ in (\ref{eq:1}) is a function of a
Chi-square random variable:
\begin{equation}
\nu = \sqrt{\frac{d}{\chi^2_d}};
\end{equation}
The degree-of-freedom $d$ of this multivariate Student-T statistic
is set to 3. The dimensionality $p$ is chosen to be 100 and we let
$\bbsg$ be the covariance matrix of an AR(1) process,
\begin{equation}
\bbsg(i,j) = r^{|i-j|}, \label{eq:cov_shape}
\end{equation}
where $\bbsg(i,j)$ denotes the entry of $\bbsg$ in row $i$ and
column $j$, and $r$ is set to 0.7 in this simulation. The sample
size $n$ varies from 5 to 225 with step size 10. All the
simulations are repeated for 100 trials and the average empirical
performance is reported.

\begin{figure}[htbp]
\centering
\includegraphics[width=.5\textwidth]{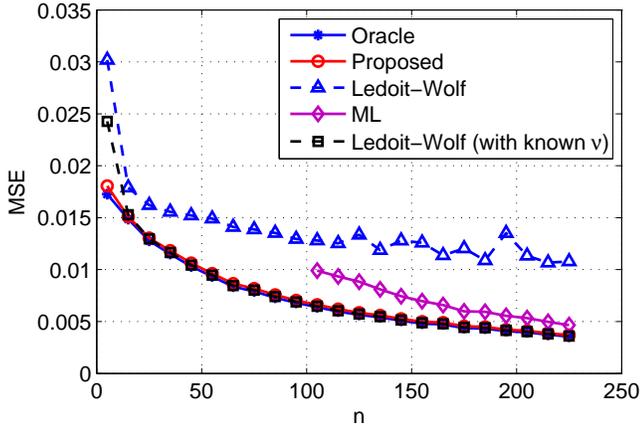}
\caption{Multivariate Student-T samples: Comparison of different
trace-normalized covariance estimators when $p = 100$.}
\label{fig:student}
\end{figure}

We use (\ref{eq:plug_in}) with $\hm = \hr$ and employ iterations
defined by (\ref{eq:iterations}) {and (\ref{eq:normalize})} with
$\rho = \hat\rho$. For comparison, we also plot the results of the
trace-normalized oracle in (\ref{eq:cov_oracle}), the
trace-normalized Ledoit-Wolf estimator \cite{Ledoit2004}, and the
non-regularized solution in (\ref{eq:7}) (when $n>p$). As the
Ledoit-Wolf estimator operates on the sample covariance which is
sensitive to outliers, we also compare to a trace-normalized
version of a clairvoyant Ledoit-Wolf estimator implemented
according to the procedure in \cite{Ledoit2004} with known $\nu$.
More specifically, the samples $\bx_i$ are firstly normalized by
the known realizations $\nu_i$, yielding truly Gaussian samples;
then the sample covariance of the normalized $\bx_i$'s is
computed, which is used to estimate the Ledoit-Wolf shrinkage
parameters and estimate the covariance via equation (14) in
\cite{Ledoit2004}. The MSE are plotted in Fig. \ref{fig:student}.
It can be observed that the proposed method performs significantly
better than the Ledoit-Wolf estimator, and the performance is very
close to the ideal oracle estimator using the optimal shrinkage
parameter (\ref{eq:cov_oracle}). Even the clairvoyant Ledoit-Wolf
implemented with known $\nu_i$ does not outperform the proposed
estimator in the small sample size regime. These results
demonstrate the robustness of the proposed approach.

\begin{figure}[htbp]
\centering \subfigure[Ledoit-Wolf]{\label{fig:student:a}
\includegraphics[width=.22\textwidth]{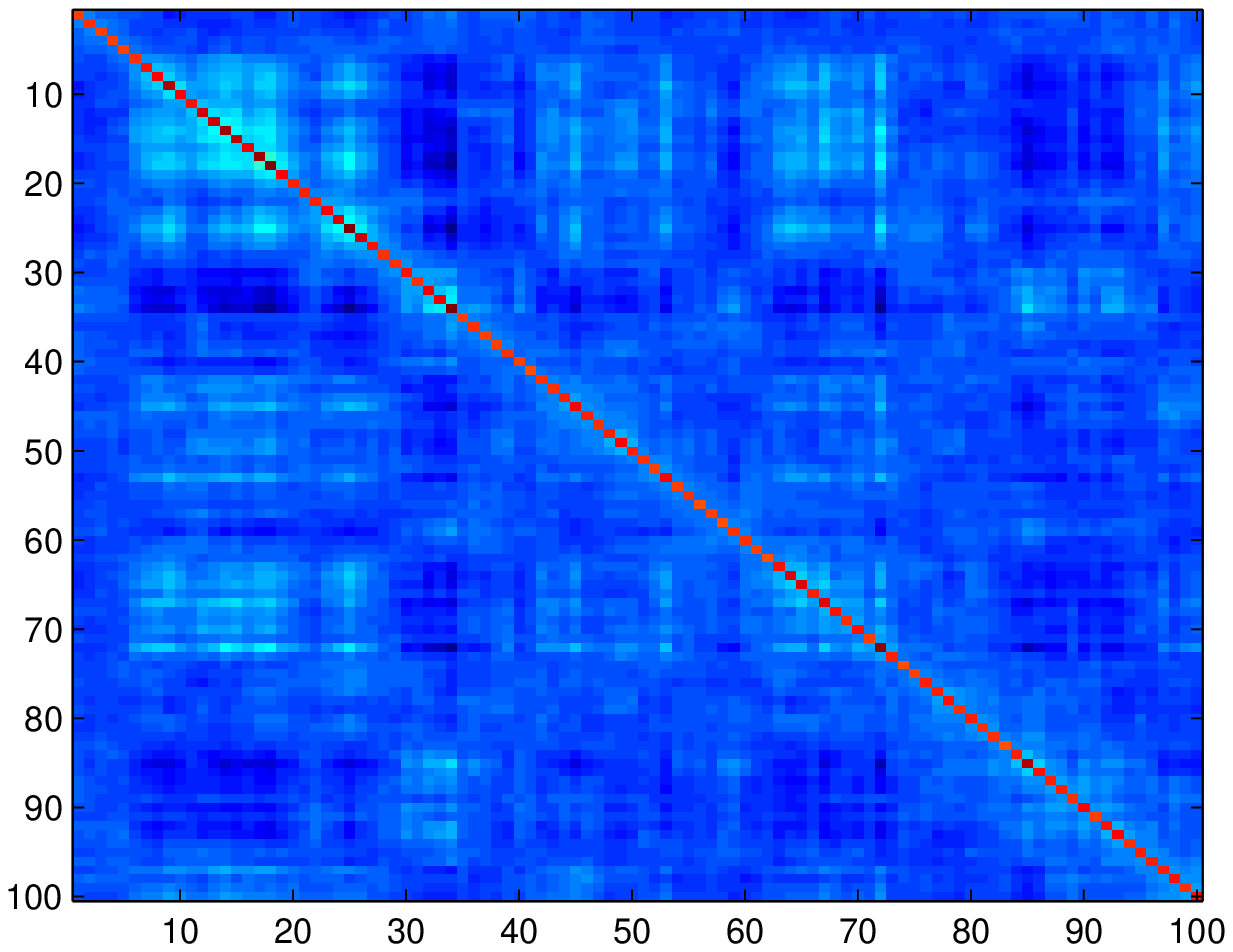}}
\subfigure[Proposed]{\label{fig:student:b}
\includegraphics[width=.22\textwidth]{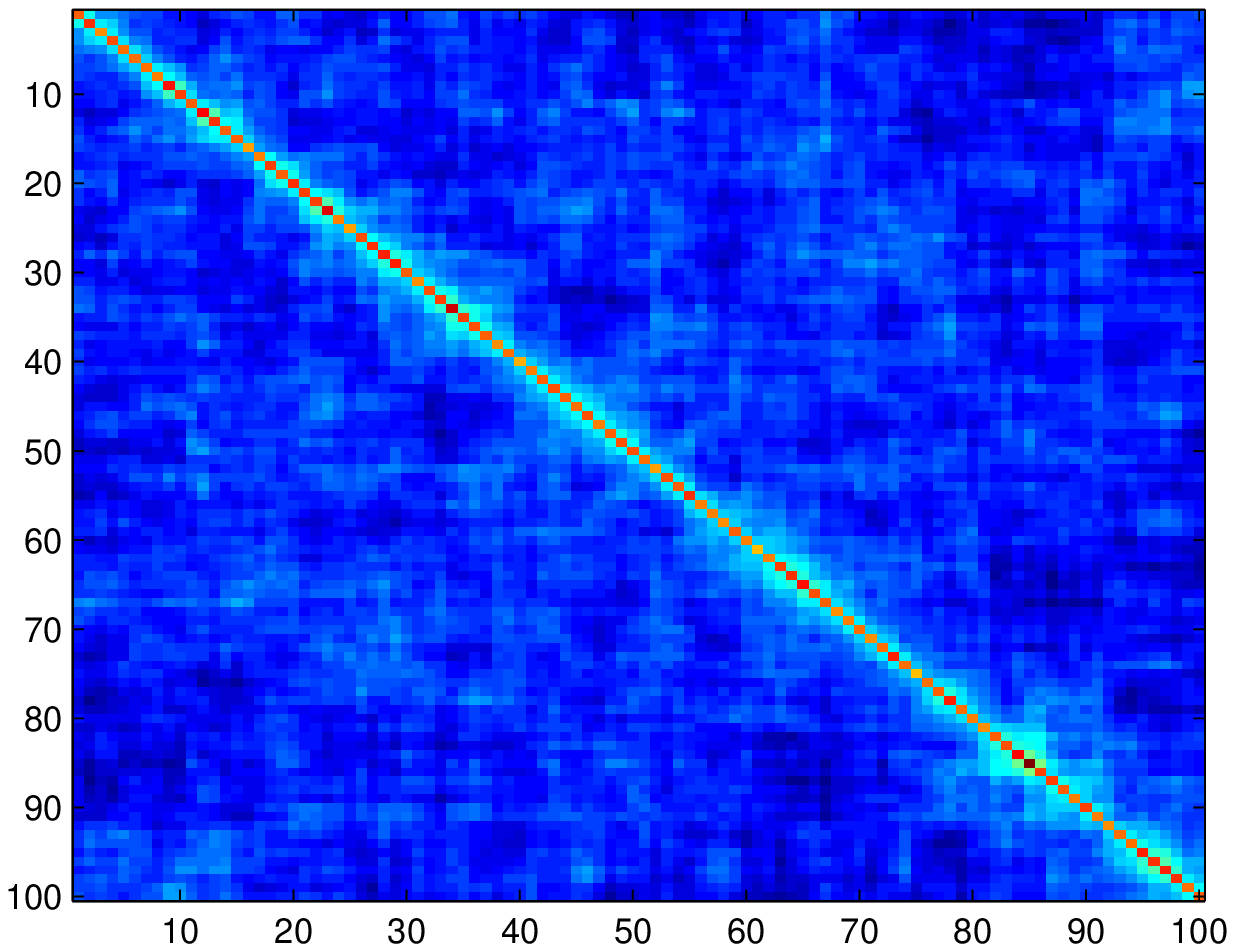}}
\caption{Multivariate Student-T samples: Visualizations of two
estimates using the Ledoit-Wolf and the proposed approaches. $p =
100, n = 50$. Note that $n<p$ in this case.}
\label{fig:comp_shape}
\end{figure}

As a graphical illustration, in Fig. \ref{fig:comp_shape} we
provide covariance visualizations for a realization of the
estimated covariances using the Ledoit-Wolf method and the
proposed approach. The sample size in this example is set to 50,
which is smaller than the dimension 100. Compared to the true
covariance, it is clear that the proposed covariance estimator
preserves the structure of the true covariance, while in the
Ledoit-Wolf covariance procudure produces ``block pattern"
artifacts caused by heavy-tails of the multivariate Student-T.

When $n>p$, we also observe a substantial improvement of the
proposed method over the ML covariance estimate, which provides
further evidence of the power of Steinian shrinkage for reducing
MSE.

\begin{figure}[htbp]
\centering
\includegraphics[width=.5\textwidth]{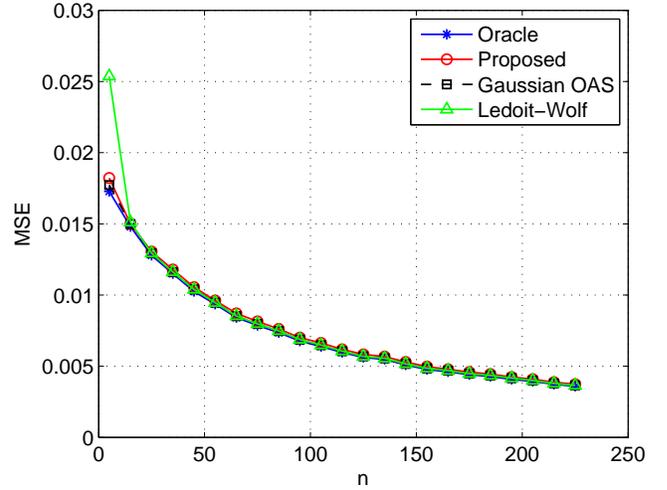}
\caption{Gaussian samples: Comparison of trace-normalized
different covariance estimators when $p = 100$.}
\label{fig:gaussian}
\end{figure}

In order to assess the tradeoff between accuracy and robustness we
investigate the case when the samples are truly Gaussian
distributed. We use the same simulation parameters as in the
previous example, the only difference being that the samples are
now generated from a Gaussian distribution. The performance
comparison is shown in Fig. \ref{fig:gaussian}, where four
different (trace-normalized) methods are included: the oracle
estimator derived from Gaussian assumptions (Gaussian oracle)
\cite{Chen}, the iterative approximation of the Gaussian oracle
(Gaussian OAS) proposed in \cite{Chen}, the Ledoit-Wolf estimator
and the proposed method. It can be seen that for truly Gaussian
samples the proposed method performs very closely to the Gaussian
OAS, which is specifically designed for Gaussian distributions.
Indeed, for small sample size ($n< 20$), the proposed method
performs even better than the Ledoit-Wolf estimator. This
indicates that, although the proposed robust method is developed
for the entire elliptical family, it actually sacrifices very
little performance for the case that the distribution is Gaussian.

\section{Application to anomaly detection in wireless sensor
networks} In this section we demonstrate the proposed robust
covariance estimator in a real application: activity detection
using a wireless sensor network.

\begin{figure}[htbp]
\centering
\includegraphics[width=.5\textwidth]{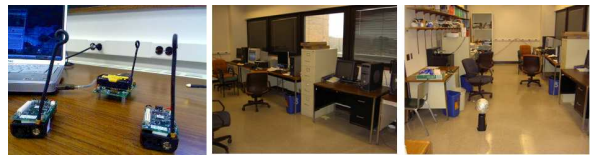}
\caption{Experimental platform: wireless Mica2 sensor nodes.}
\label{fig:Mica2}
\end{figure}

The experiment was set up on an Mica2 sensor network platform, as
shown in Fig. \ref{fig:Mica2}, which consists of 14 sensor nodes
randomly deployed inside and outside a laboratory at the
University of Michigan. Wireless sensors communicated with each
other asynchronously by broadcasting an RF signal every 0.5
seconds. The received signal strength (RSS), defined as the
voltage measured by a receiver's received signal strength
indicator circuit (RSSI), was recorded for each pair of
transmitting and receiving nodes. There were $14 \times 13 = 182$
pairs of RSSI measurements over a 30 minute period, and samples
were acquired every 0.5 sec. During the experiment period, persons
walked into and out of the lab at random times, causing anomaly
patterns in the RSSI measurements. Finally, for ground truth, a
web camera was employed to record the actual activity.

\begin{figure}[htbp]
\centering
\includegraphics[width=.5\textwidth]{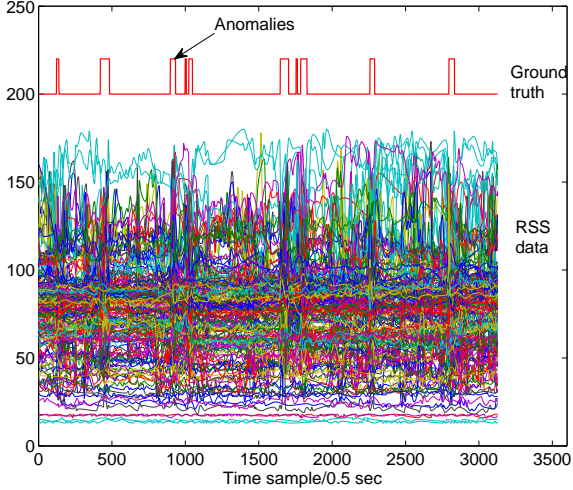}
\caption{At bottom 182 RSS sequences sampled from each pair of
transmitting and receiving nodes in intrusion detection
experiment. Ground truth indicators at top are extracted from
video captured from a web camera that recorded the scene.}
\label{fig:rssdata}
\end{figure}

\begin{figure}[htbp]
\centering
\includegraphics[width=.5\textwidth]{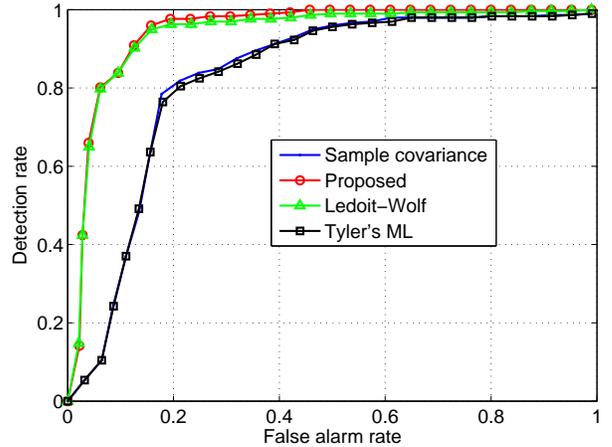}
\caption{Performance comparison for different covariance
estimators, $p = 182, n = 200$. The covariance matrix is estimated
in a \emph{supervised} manner.} \label{fig:roc_clearn}
\end{figure}

Fig. \ref{fig:rssdata} shows all the received signals and the
ground truth indicator extracted from the video. The objective of
this experiment was intrusion (anomaly) detection. We emphasize
that, with the exception of the results shown in Fig.
\ref{fig:roc_clearn},  the ground truth indicator is only used for
performance evaluation and the detection algorithms presented here
were completely \emph{unsupervised}.

To remove temperature drifts \cite{AnalogIC} of receivers we
detrended the data as follows. Let $x_i[k]$ be the $k$-th sample
of the $i$-th RSS signal and denote
\begin{equation}
\bx[k] = \bl x_1[k], x_2[k], \ldots, x_{182}[k]\br^T.
\end{equation}
The local mean value of $\bx[k]$ is defined by
\begin{equation}
\bar{\bx}[k] = \frac{1}{2m+1}\sum_{i=k-m}^{k+m} \bx[k],
\end{equation}
where the integer $m$ determines local window size and is set to
50 in this study. We detrend the data by subtracting this local
mean
\begin{equation}
\by[k] = \bx[k]-\bar{\bx}[k],
\end{equation}
yielding a detrended sample $\by[k]$ used in our anomaly
detection.

\begin{figure}[htbp]
\centering
\includegraphics[width=.5\textwidth]{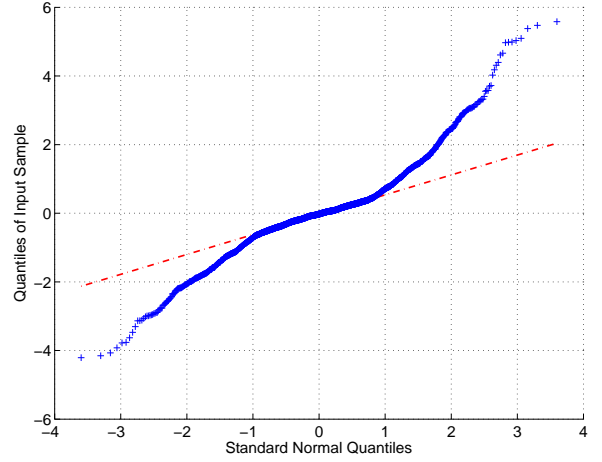}
\caption{QQ plot of data versus the standard Gaussian
distribution.} \label{fig:qqplot}
\end{figure}

\begin{figure}[htbp]
\centering
\includegraphics[width=.5\textwidth]{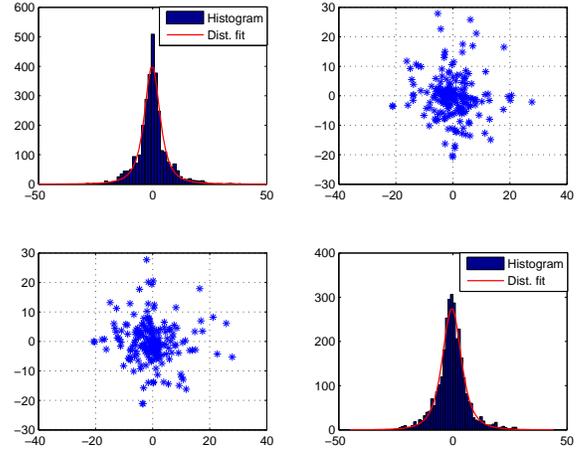}
\caption{Histograms and scatter plots of the first two de-trended
RSS sequences, which are fit by a multivariate Student-T
distribution with degree-of-freedom $d = 2$.}
\label{fig:scatterplot}
\end{figure}

We established that the detrended measurements were heavy-tailed
non-Gaussian by performing several statistical tests. First the
Lilliefors test \cite{Lilliefors} of Gaussianity was performed on
the detrended RSS measurements. This resulted in rejection of the
Gaussian hypothesis at a level of significance of $10^{-6}$. As
visual evidence, we show the quantile-quantile plot (QQ plot) for
one of the detrended RSS sequences in Fig. \ref{fig:qqplot} which
illustrates that the samples are non-Gaussian. In Fig.
\ref{fig:scatterplot}, we plot the histograms and scatter plots of
two of the detrended RSS sequences, which shows the heavy-tail
nature of the sample distribution. This strongly suggests that the
RSS samples can be better described by a heavy-tailed elliptical
distribution than by a Gaussian distribution. As additional
evidence, we fitted a Student-T distribution to the first
detrended RSS sequence, and used maximum likelihood to estimate
the degree-of-freedom as $d = 2$ with a $95\%$ confidence interval
(CI) $[1.8460, 2.2879]$.

\begin{figure}[htbp]
\centering
\includegraphics[width=.5\textwidth]{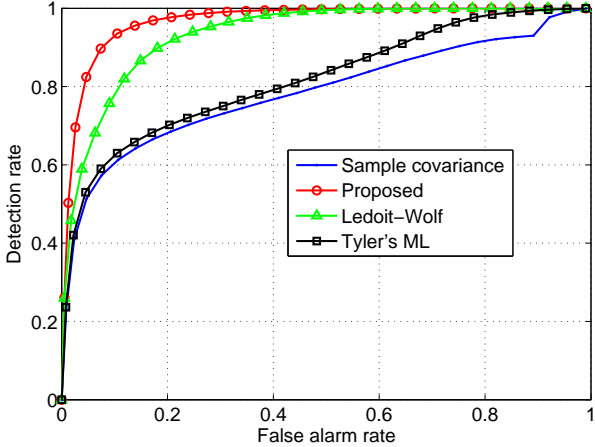}
\caption{Performance comparison for different covariance
estimators, $p = 182, n = 200$.} \label{fig:roc}
\end{figure}

Consider the following function of the detrended data:
\begin{equation}
    \label{eq:test_stats}
t_k = \by[k]^T \bbsg^{-1} \by[k].
\end{equation}
for known $\bbsg = E\left\{ \mathbf y[k] \mathbf y [k]^T
\right\}$, $t_k$ is a statistic that has been previously proposed
for anomaly detection \cite{AnomalyDetectionSurvey}. A time sample
is declared to be anomalous if the test statistic $t_k$ exceeds a
specified threshold. We then applied our proposed robust
covariance estimator to estimate the unknown $\bbsg$ and
implemented (\ref{eq:test_stats}) for activity detection.
Specifically, we constructed the $182 \times 182$ sample
covariance by randomly subsampling 200 time slices from the RSS
data shown in Fig. \ref{fig:rssdata}. Note, that these 200 samples
correspond to a training set that is contaminated by  anomalies at
the same anomaly rate (approximately 10\%) as the entire sample
set. The detection performance was evaluated using the receiver
operating characteristic (ROC) curve, where the averaged curves
from 200 independent Monte-Carlo trials are shown in Fig.
\ref{fig:roc}. For comparison, we also implemented the activity
detector (\ref{eq:test_stats}) with other covariance estimates
including: the sample covariance, the Ledoit-Wolf estimator and
Tyler's ML estimator.

From the mean ROCs we can see that the detection performances are
rank ordered as follows: Proposed $>$ Ledoit-Wolf $>$ Tyler's ML
$>$ Sample covariance. The sample covariance performs poorly in
this setting due to the small sample size ($n = 200, p = 182$) and
its sensitivity to the heavy-tailed distribution shown in Fig.
\ref{fig:qqplot} and \ref{fig:scatterplot}. The Tyler ML method
and the Ledoit-Wolf estimator improve upon the sample covariance
since they compensate for heavy tails and for small sample size,
respectively. Our proposed method compensates for both effects
simultaneously and achieves the best detection performance.

\begin{figure}[htbp]
\centering
\includegraphics[width=.5\textwidth]{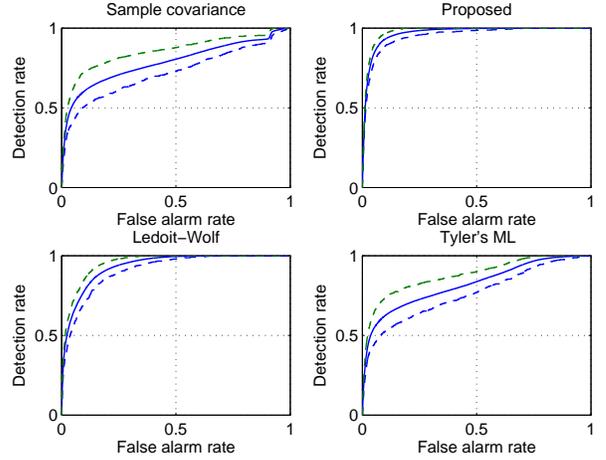}
\caption{Performance comparison for different covariance
estimators, including the mean value and 90$\%$ confidence
intervals. (a): Sample covariance. (b): Proposed. (c):
Ledoit-Wolf. (d): Tyler's ML. The 200 training samples are
randomly selected from the entire data set.} \label{fig:four_rocs}
\end{figure}

We also plot the 90$\%$ confidence envelopes, determined by
cross-validation, on the ROCs in Fig. \ref{fig:four_rocs}. The
width of the confidence interval reflects the sensitivity of the
anomaly detector to variations in the training set. Indeed, the
upper and lower endpoints of the confidence interval are the
optimistic and the pessimistic predictions of detection
performance. The proposed method achieves the smallest width among
the four computed $90\%$ confidence envelopes.

Finally, for completeness we provide performance comparison of
covariance-based \emph{supervised} activity detection algorithms
in Fig. \ref{fig:roc_clearn}. The training period is selected to
be $[251,450]$ based on ground truth where no anomalies appear. It
can be observed that, by excluding the outliers caused by
anomalies, the performance of the Ledoit-Wolf based intrusion
detection algorithm is close to that of the proposed method. We
conclude that the activity detection performance of the proposed
covariance estimator is more robust than the other three
estimators with respect to outlier contamination in the training
samples.

\section{Conclusion}
In this paper, we proposed a shrinkage covariance estimator which
is robust over the class of elliptically distributed samples. The
proposed estimator is obtained by fixed point iterations, and we
established theoretical guarantees for existence, convergence and
uniqueness. The optimal shrinkage coefficient was derived using a
minimum mean-squared-error framework and has a closed-form
expression in terms of the unknown true covariance. This
expression can be well approximated by a simple plug-in estimator.
Simulations suggest that the iterative approach converges to a
limit which is robust to heavy-tailed multivariate Student-T
samples. Furthermore, we show that for the Gaussian case, the
proposed estimator performs very closely to previous estimators
designed expressly for Gaussian samples.

As a real world application we demonstrated the performance of the
proposed estimator in intrusion detection using a wireless sensor
network. Implementation of a standard covariance-based detection
algorithm using our robust covariance estimator achieved superior
performances as compared to conventional covariance estimators.

The basis of the proposed method is the ML estimator originally
proposed by Tyler in \cite{Tyler87}. However, the approach
presented in this paper can be extended to other M-estimators.

One of the main contributions of our work is the proof of
uniqueness and convergence of the estimator. This proof extends
the results of \cite{Tyler87,Pascal} to the regularized case.
Recently, an alternative proof to the non-regularized case using
convexity on manifolds was presented in \cite{Auderset}. This
latter proof highlights the geometrical structure of the problem
and gives additional insight. We are currently investigating its
extension to the regularized case.

\section{Acknowledgement}
The authors would like to thank Neal Patwari for designing the
intrusion detection experiment and collecting the data. We also
thank Harry Schmidt and his group at Raytheon, Tucson AZ, for
lending us the Mica2 motes for this experiment. We would like to
thank Prof. Drton for fruitful discussions and a preprint of
\cite{Drton}.

\section{Appendix}
\subsection{Proof of Theorem \ref{thm:2}}
In this appendix we prove Theorem \ref{thm:2}. The original
convergence proof for the non-regularized case in
\cite{Tyler87,Pascal} is based on careful exploitation of the
specific form of (\ref{eq:4}). In contrast, our proof for the
regularized case is based on a direct connection from the concave
Perron-Frobenius theory \cite{Krause94, Krause01} that is simpler
and easier to generalize. We begin by summarizing the required
concave Perron-Frobenius result in the following lemma.

\begin{lem}[\cite{Krause94}]
    \label{lem:concavePerronF}
Let $(E, \|\cdot\|)$  be a Banach space with $K \subset E$ being a
closed, convex cone on which $\|\cdot \|$ is increasing, i.e., for
which $x \le y$ implies $\|x\|\le \|y\|$, where the operator $\le$
on the convex cone $K$ means that if $x \le y$ then $y-x \in K$.
Define $U = \left\{ {x|x \in K,||x|| = 1} \right\}$. Let $T:K \to
K$ be a concave operator such that
\begin{equation}
    \label{eq:concave}
\begin{aligned}
& T(\mu x + (1-\mu) y) \ge \mu T(x) + (1-\mu) T(y), \\
& \quad \quad \text{for all } \mu \in [0,1], ~~ \text{all } x,y
\in K.
\end{aligned}
\end{equation}
If for some $e \in K-\{0\}$ and constants $a>0$, $b>0$ there is
\begin{equation}
    \label{eq:proveab}
ae \le T(x) \le be, \quad \text{for all }x\in U,
\end{equation}
then there exists a unique $x^* \in U$ to which the iteration of
the normalized operator {$\tilde{T}(x) = T(x)/\|T(x)\|, x\in K-
\{0\}$} converges:
\begin{equation}
    \label{eq:lemma_converge}
\mathop {\lim }\limits_{k \to \infty } \tilde T^k (x) = x^*, ~~
\text{for all } x \in K -\{0\}.
\end{equation}
\end{lem}
Lemma \ref{lem:concavePerronF} can be obtained by combining
results from {Lemma 2 and Theorem} in Section 4 of
\cite{Krause94}. Here we show that the proof of Theorem
\ref{thm:2} is a direct result of applying Lemma
\ref{lem:concavePerronF} with proper definitions of $E$, $K$, $U$
and $T$:

\begin{itemize}
  \item $E$: the set of all symmetric matrices;
  \item $K$: the set of all positive semi-definite
matrices;
  \item $\|\bbsg\|$: the normalized nuclear norm of $\bbsg$, \ie
  \begin{equation}
\|\bbsg\| = \frac{1}{p} \sum_{j=1}^p |\lambda_j|,
\end{equation}
where $\lambda_j$ is the $j$-th eigenvalue of $\bbsg$ and
$|\cdot|$ is the absolute value operator. {Note that for any
$\bbsg \in K$, the nuclear norm $\|\cdot\|$ is identical to
$\tr(\cdot)/p$ and is increasing};
  \item $U$: the set $U = \{\bbsg | \bbsg \in K, \|\bbsg\| =1\}$;
  \item $T$: the mapping from $K$ to $K$ defined by
  \begin{equation}
    \label{eq:mapping}
  T(\bbsg) = (1-\rho)\frac{p}{n}\sum_{i=1}^n w(\bs_i,\bbsg)
  \bs_i\bs_i^T + \rho \bbi,
  \end{equation}
  where the weight function $w(\bs_i,\bbsg)$ is defined as
  \begin{equation}
  w(\bs_i,\bbsg) = \inf_{\bz^T\bs_i \ne 0}
  \frac{\bz^T\bbsg\bz}{(\bs_i^T\bz)^2},
  \end{equation}
for any $\bbsg\in K$.
\end{itemize}

\IEEEproof With the above definitions we show that Theorem
\ref{thm:2} is a direct result of Lemma \ref{lem:concavePerronF}.
We begin by showing that the mapping operator $T$ is concave.
Indeed, it is sufficient to show that $w(\bs_i,\bbsg)$ in concave
in $\bbsg$, {which is true because it is the infinimum of affine
functions of $\bbsg$}.
%

Next, we show that $T$ satisfies condition (\ref{eq:proveab}) with
$e = \bbi$. It is easy to see that
\begin{equation}
\rho \bbi \le T(\bbsg),
\end{equation}
for any $\bbsg \in U$. Then we show that
\begin{equation}
    \label{eq:w}
w(\bs_i,\bbsg) \le p,
\end{equation}
for any $\bbsg \in U$. Indeed,
\begin{equation}
    \label{eq:boundw}
w(\bs_i,\bbsg) =\inf_{\bz^T\bs_i \ne
0}\frac{\bz^T\bbsg\bz}{(\bs_i^T\bz)^2} \le
\frac{\bs_i^T\bbsg\bs_i}{(\bs_i^T\bs_i)^2} \le
\frac{\lambda_{\max}}{\bs_i^T\bs_i} = \lambda_{\max},
\end{equation}
where $\lambda_{\max}$ is the maximum eigenvalue of $\bbsg$. The
last equality in the right-hand-side of (\ref{eq:boundw}) comes
from the fact that $\bs_i$ is of unit norm by definition
(\ref{eq:5}). (\ref{eq:w}) is thus obtained by noticing that
$\bbsg \in U$ and $\lambda_{\max} \le p$. Substituting
(\ref{eq:w}) into (\ref{eq:mapping}) we have
\begin{equation}
T(\bbsg) \le (1-\rho) p^2 \hr + \rho\bbi \le
\bl(1-\rho)p^2\alpha_{\max} + \rho\br \bbi,
\end{equation}
where
\[\hr = \frac{1}{n}\sum_{i=1}^n \bs_i\bs_i^T,\] and
$\alpha_{\max}$ is the maximum eigenvalue of $\hr$. Again, as
$\bs_i$ is of unit norm, $\alpha_{\max} \le \tr(\hr) = 1$ and
\begin{equation}
T(\bbsg) \le \bl (1-\rho)p^2 + \rho\br \bbi.
\end{equation}
Therefore, we have shown that $T$ satisfies condition
(\ref{eq:proveab}), where $e = \bbi$, $a = \rho$ and $b =
(1-\rho)p^2 + \rho$. In addition, (\ref{eq:proveab}) establishes
that the mapping $T$ from $U$ always yields a positive definite
matrix. Therefore, the convergent limit of the fixed-point
iteration is positive definite.

Finally, we note that, for any $\bbsg \succ \zero$, we have
\begin{equation}
\|\bbsg\| = \frac{\tr(\bbsg)}{p},
\end{equation}
and
\begin{equation}
w(\bs_i,\bbsg) = \inf_{\bz^T\bs_i \ne
0}\frac{\bz^T\bbsg\bz}{(\bs_i^T\bz)^2} =
\frac{1}{\bs_i^T\bbsg^{-1}\bs_i}.
\end{equation}
The limit (\ref{eq:lemma_converge}) is then identical to the limit
of proposed iterations (\ref{eq:iterations}) and
(\ref{eq:normalize}) for any $\bbsg \succ \zero$. Therefore,
Theorem \ref{thm:2} has been proved.

\subsection{Proof of Theorem \ref{thm:rho}}
\IEEEproof To ease the notation we define  $\tc$ as
\begin{equation}
\tc =  \frac{p}{n}\sum_{i=1}^n \frac{\bs_i \bs_i^T}{\bs_i^T
\bbsg^{-1} \bs_i}.
\end{equation}
The shrinkage estimator in (\ref{eq:tsg_def}) is then
\begin{equation}
    \label{eq:hsghc_def}
\tsg(\rho) = (1-\rho) \tc + \rho \bbi.
\end{equation}
By substituting (\ref{eq:hsghc_def}) into (\ref{eq:tsg_def}) and
taking derivatives of $\rho$, we obtain that
\begin{equation}
    \label{eq:rhoO}
\begin{aligned}
\rho_O & = \frac{E\blc\tr\bl(\bbi-\tc)(\bbsg-\tc)\br\brc}{E\blc\l\|\bbi-\tc\r\|_F^2\brc}\\
& = \frac{m_2-m_{11}-m_{12}+{\tr(\bbsg)}}{m_2-2m_{11}+p},
\end{aligned}
\end{equation}
where
\begin{equation}
m_2 = E\blc\tr(\tc^2)\brc,
\end{equation}
\begin{equation}
m_{11} = E\blc\tr(\tc)\brc,
\end{equation}
and
\begin{equation}
m_{12} = E\blc\tr(\tc\bbsg)\brc.
\end{equation}

Next, we calculate the moments. We begin by eigen-decomposing
$\bbsg$ as
\begin{equation}
\bbsg = \bbu \bbd \bbu^T,
\end{equation}
and denote
\begin{equation}
\bbla = \bbu \bbd^{1/2}.
\end{equation}
Then, we define
\begin{equation}
\bz_i = \frac{\bbla^{-1}  \bs_i}{\|\bbla^{-1} \bs_i\|_2} =
\frac{\bbla^{-1} \bu_i}{\|\bbla^{-1} \bu_i\|_2}.
\end{equation}
Noting that $\bu_i$ is a Gaussian distributed random vector with
covariance $\bbsg$, it is easy to see that $\|\bz_i\|_2=1$ and
$\bz_i$ and $\bz_j$ are independent with each other for $i \ne j$.
Furthermore, $\bz_i$ is isotropically distributed\cite{Marzetta,
Hassibi, Eldar_} and satisfies \cite{Chen, randommatrixbook}
\begin{equation}
    \label{eq:Ezz}
E\blc \bz_i \bz_i^T\brc = \frac{1}{p} \bbi,
\end{equation}
\begin{equation}
\label{eq:ziDzi2}
\begin{aligned}
E\blc \bl \bz_i^T \bbd \bz_i \br^2\brc & = \frac{1}{p(p+2)}\bl2\tr(\bbd^2) + \tr^2(\bbd)\br\\
& =  \frac{1}{p(p+2)}\bl2\tr(\bbsg^2) + \tr^2(\bbsg)\br,
\end{aligned}
\end{equation}
and
\begin{equation}
    \label{eq:EziDzj2}
E\blc\bl \bz_i^T \bbd \bz_j \br^2\brc = \frac{1}{p^2} \tr(\bbd^2)
=  \frac{1}{p^2} \tr(\bbsg^2), ~i \ne j.
\end{equation}

Expressing $\tc$ in terms of $\bz_i$, there is
\begin{equation}
    \label{eq:hcz}
\tc = \frac{p}{n} \bbla \sum_{i=1}^n \bz_i \bz_i^T \bbla^T.
\end{equation}
Then,
\begin{equation}
E\blc\tc\brc =  \frac{p}{n} \bbla \sum_{i=1}^n E\blc \bz_i
\bz_i^T\brc \bbla^T = \bbsg,
\end{equation}
and accordingly we have
\begin{equation}
    \label{eq:m11}
\begin{aligned}
m_{11} = E\blc\tr(\tc)\brc = \tr(\bbsg) ,
\end{aligned}
\end{equation}
and
\begin{equation}
    \label{eq:m12}
\begin{aligned}
m_{12} = E\blc\tr(\tc\bbsg)\brc = \tr(\bbsg^2).
\end{aligned}
\end{equation}
For $m_2$ there is
\begin{equation}
    \label{eq:m2}
\begin{aligned}
m_2 & = \frac{p^2}{n^2} E\blc\tr\bl\bbla \sum_{i=1}^n \bz_i \bz_i^T \bbla^T \bbla \sum_{j=1}^n \bz_j \bz_j^T \bbla^T \br\brc\\
& = \frac{p^2}{n^2} E\blc\tr\bl \sum_{i=1}^n \sum_{j=1}^n \bz_i \bz_i^T \bbla^T \bbla  \bz_j \bz_j^T \bbla^T \bbla\br\brc\\
& =  \frac{p^2}{n^2} E\blc\tr\bl \sum_{i=1}^n \sum_{j=1}^n \bz_i \bz_i^T \bbd  \bz_j \bz_j^T \bbd\br\brc\\
& =  \frac{p^2}{n^2} \sum_{i=1}^n \sum_{j=1}^n E\blc \bl \bz_i^T
\bbd  \bz_j \br^2\brc.
\end{aligned}
\end{equation}
Now substitute (\ref{eq:ziDzi2}) and (\ref{eq:EziDzj2}) to
(\ref{eq:m2}): \small
\begin{equation}
    \label{eq:m2_}
\begin{aligned}
m_2 & = \frac{p^2}{n^2} \bl \frac{n}{p(p+2)}\bl2\tr(\bbsg^2) + \tr^2(\bbsg)\br + \frac{n(n-1)}{p^2} \tr(\bbsg^2) \br\\
& = \frac{1}{n(1+2/p)} \bl2\tr(\bbsg^2) + \tr^2(\bbsg)\br + (1-\frac{1}{n})\tr(\bbsg^2)\\
& = \bl 1 - \frac{1}{n} + \frac{2}{n(1+2/p)}\br \tr(\bbsg^2) +
\frac{\tr^2(\bbsg)}{n(1+2/p)}.
\end{aligned}
\end{equation}
\normalsize

Recalling $\tr(\bbsg) = p$, (\ref{eq:cov_oracle}) is finally
obtained by substituting (\ref{eq:m11}), (\ref{eq:m12}) and
(\ref{eq:m2_}) into (\ref{eq:rhoO}).

\end{document}

%% file: Defn.tex
\newcommand{\bl}{\left(}
\newcommand{\br}{\right)}
\newcommand{\blc}{\left\{}
\newcommand{\brc}{\right\}}

\newcommand{\bs}{{\bf s}}

\newcommand{\by}{{\bf y}}

\newcommand{\bx}{{\bf x}}

\newcommand{\bu}{{\bf u}}
\newcommand{\bz}{{\bf z}}
\newcommand{\bbsg}{{\bf \Sigma}}
\newcommand{\bbla}{\mathbf \Lambda}
\newcommand{\bbi}{{\bf I}}

\newcommand{\zero}{{\bf 0}}

\newcommand{\bbd}{{{\bf D}}}

\newcommand{\bbbr}{{{\bf R}}}

\newcommand{\bbs}{{\bf S}}

\newcommand{\bbu}{{\bf U}}

\newcommand{\bbc}{{\bf C}}

\newcommand{\hs}{\widehat{\bbs}}
\newcommand{\hr}{\widehat{\bbbr}}

\newcommand{\hsg}{\widehat{\bbsg}}

\newcommand{\tc}{\widetilde{\bbc}}
\newcommand{\tsg}{\widetilde{\bbsg}}

\newcommand{\bbm}{{\bf M}}

\newcommand{\hm}{{\widehat{\bbm}}}

\newcommand{\ie}{{\em i.e., }}
\newcommand{\eg}{{\em e.g., }}

\newcommand{\tr}{\mathrm{Tr}}

\newtheorem{lem}{Lemma}
\newtheorem{thm}{Theorem}

\def\({\left (}
\def\){\right)}
\def\l{\left}
\def\r{\right}
\def\det{{\mathrm{det}}}